\documentclass[review]{elsarticle}
\journal{Neurocomputing}

\usepackage[utf8]{inputenc} 
\usepackage[T1]{fontenc}    
\usepackage{hyperref}       
\usepackage{url}            
\usepackage{booktabs}       
\usepackage{amsfonts}       
\usepackage{nicefrac}       
\usepackage{microtype}      
\usepackage{lipsum}
\usepackage{graphicx}
\usepackage{multirow}
\usepackage{stackengine}
\usepackage[table]{xcolor}
\usepackage{array}
\usepackage{framed,multirow}
\usepackage{amssymb} 
\usepackage{amsmath,amssymb}
\usepackage{color}
\def\btheta{\boldsymbol{\theta}}
\begin{document}
\begin{frontmatter}
\title{Interpreting chest X-rays via CNNs that exploit \\ hierarchical disease dependencies and uncertainty labels}

\author{Hieu H. Pham$^{*}$, Tung T. Le, Dat Q. Tran, Dat T. Ngo, Ha Q. Nguyen}
\address{Medical Imaging Group, Vingroup Big Data Institute (VinBDI) \\ 458 Minh Khai street, Hai Ba Trung, Hanoi, Vietnam\\[-0.6cm]}
\cortext[mycorrespondingauthor]{Corresponding author:\textcolor{blue}{\texttt{ \underline{v.hieuph4@vinbdi.org}}}  (Hieu H. Pham)}

\begin{abstract}
Chest radiography is one of the most common types of diagnostic radiology exams, which is critical for screening and diagnosis of many different thoracic diseases. Specialized algorithms have been developed to detect several specific pathologies such as lung nodules or lung cancer. However, accurately detecting the presence of multiple diseases from chest X-rays (CXRs) is still a challenging task. This paper presents a supervised multi-label classification framework based on deep convolutional neural networks (CNNs) for predicting the presence of 14 common thoracic diseases and observations. We tackle this problem by training state-of-the-art CNNs that exploit hierarchical dependencies among abnormality labels. We also propose  to use the label smoothing technique for a better handling of uncertain samples, which occupy a significant portion of many CXR datasets. Our model is trained  on over 200,000 CXRs of the recently released CheXpert dataset and achieves a mean area under the curve (AUC) of 0.940 in predicting 5 selected pathologies from the validation set. This is the highest AUC score yet reported to date. The proposed method is also evaluated on the independent test set of the CheXpert competition, which is composed of 500 CXR studies annotated by a panel of 5 experienced radiologists. The performance is on average better than 2.6 out of 3 other individual radiologists with a mean AUC of 0.930, which ranks first on the CheXpert leaderboard at the time of writing this paper. 
\end{abstract}

\end{frontmatter}
\section{Introduction}
Chest X-ray (CXR) is one of the most common radiological exams in diagnosing many different diseases related to lung and heart, with millions of scans performed globally every year~\cite{NHS_England,anderson2015global}. Many diseases among them, like \textit{Pneumothorax}~\cite{bellaviti2016increased}, can be deadly if not diagnosed quickly and accurately enough. 
A computer-aided diagnosis (CAD) system that is able to correctly diagnose the most common observations from CXRs will significantly benefit  many clinical practices. In this work, we investigate the problem of multi-label classification for CXRs using deep convolutional neural networks (CNNs).

There has been a recent effort to harness advances in machine learning, especially deep learning, to build a new generation of CAD systems for classification and localization of common thoracic diseases from CXR images~\cite{qin2018computer}. Several motivations are behind this transformation: First, interpreting CXRs to accurately diagnose pathologies is difficult. Even well-trained radiologists can easily make mistake due to challenges in distinguishing different kinds of pathologies, many of which often have similar visual features~\cite{delrue2011difficulties}. Therefore, a high-precision method for common thorax diseases classification and localization can be used as a second reader to support the decision making process of radiologists and to help reduce the diagnostic error. It also addresses the lack of diagnostic expertise in areas where the radiologists are limited or not available~\cite{crisp2014global,theatlantic}. Second, such a system can be used as a screening tool that helps reduce waiting time of patients in hospitals and allows care providers to respond to emergency situations sooner or to speed up a diagnostic imaging workflow~\cite{annarumma2019automated}. Third, deep neural networks, in particular deep CNNs, have shown remarkable performance for various applications in medical imaging analysis~\cite{Litjens-et-al:2017}, including the CXR interpretation task~\cite{rajpurkar2017chexnet,guan2018diagnose,rajpurkar2018deep,kumar2018boosted}. 

Several deep learning-based approaches have been proposed for classifying lung diseases and proven that they could achieve human-level performance~\cite{rajpurkar2017chexnet,hwang2019development}. Almost all of these approaches, however, aim to detect some specific diseases such as pneumonia~\cite{jaiswal2019identifying}, tuberculosis~\cite{lakhani2017deep,pasa2019efficient}, or lung cancer~\cite{ausawalaithong2018automatic}. Meanwhile, building a unified deep learning framework for accurately detecting the presence of multiple common thoracic diseases from CXRs remains a difficult task that requires much research effort. In particular, we recognize that standard multi-label classifiers often ignore domain knowledge. For example, in the case of CXR data, how to leverage clinical taxonomies of disease patterns and how to handle uncertainty labels are still open questions, which have not received much research attention. This observation motivates us to build and optimize a predictive model based on deep CNNs for the CXR interpretation in which dependencies among labels and uncertainty information are taken into account during both the training and inference stages. Specifically, we develop a deep learning-based approach that puts together the ideas of \emph{conditional training}~\cite{Chen2019DeepHM} and \emph{label smoothing}~\cite{Muller2019WhenDL} into a novel training procedure for classifying 14 common lung diseases and observations. We trained our system on more than 200,000 CXRs of the CheXpert dataset~\cite{Irvin2019CheXpertAL}---one of the largest CXR dataset currently available, and evaluated it on the validation set of CheXpert containing 200 studies, which were manually annotated by 3 board-certified radiologists. The proposed method is also tested against the majority vote of 5 radiologists on the hidden test set of the CheXpert competition that contains 500 studies.


This study makes several contributions. First, we propose a novel training strategy for multi-label CXR classification that incorporates (1) a conditional training process based on a predefined disease hierarchy and (2) a smoothing regularization technique for uncertainty labels. The benefits of these two key factors are empirically demonstrated through our ablation studies. Second, we train a series of state-of-the-art CNNs on frontal-view CXRs of the CheXpert dataset for classifying 14 common thoracic diseases. Our best model, which is an ensemble of various CNN architectures, achieves the highest area under ROC curve (AUC) score on both the validation set and test set of CheXpert at the time being. Specifically, on the validation set, it yields an averaged AUC of 0.940 in predicting 5 selected lung diseases: \textit{Atelectasis} (0.909), \textit{Cardiomegaly} (0.910), \textit{Edema} (0.958), \textit{Consolidation} (0.957) and \textit{Pleural Effusion} (0.964). This model improves the baseline method reported  in~\cite{Irvin2019CheXpertAL} by a large margin of 5\%. On the independent test set, we obtain a mean AUC of 0.930. More importantly, the proposed deep learning model is on average more accurate than 2.6 out of 3 individual radiologists in predicting the 5 selected thoracic diseases when presented with the same data\footnote{Our model (Hierarchical-Learning-V1) currently takes the first place in the CheXpert competition. More information can be found at \url{https://stanfordmlgroup.github.io/competitions/chexpert/}. Updated on \today}.

The rest of the paper is organized as follows. Related works on CNNs in medical imaging and the problem of multi-label classification in CXR images are reviewed in Section~\ref{sec:2}. In Section~\ref{sec:3}, we present the details of the proposed method with a focus on how to deal with dependencies among diseases and uncertainty labels. Section~\ref{sec:4} provides comprehensive experiments on the CheXpert dataset. Section~\ref{sec:5} discusses the experimental results, some key findings and limitations of this research. Finally, Section~\ref{sec:6} concludes the paper.

\section{Related works}
\label{sec:2}

Thanks to the increased availability of large scale, high-quality labeled datasets \cite{wang2017chestx,Irvin2019CheXpertAL,johnson2019mimic} and high-performing deep network architectures ~\cite{he2016deep,huang2017densely,szegedy2017inception,zoph2018learning}, deep learning-based approaches have been able to reach, even outperform expert-level performance for many medical image interpretation tasks \cite{rajpurkar2017chexnet,rajpurkar2018deep,guan2018diagnose,shen2019deep,huang2017added,lakhani2017deep}. Most successful applications of deep neural networks in medical imaging rely on CNNs~\cite{lecun1998gradient,NIPS2012_4824}, which utilize convolutions to extract local features of the medical images.

For CXR interpretation, the multi-label classification is a common setting in which each training example is associated with possibly more than one label~\cite{zhang2013review,tsoumakas2007multi}. Due to its important role in medical imaging, a variety of approaches have been proposed in the literature. For instance, Rajpurkar et \textit{al}.~\cite{rajpurkar2017chexnet} introduced CheXNet---a DenseNet-121 model that was trained on the ChestX-ray14 dataset~\cite{wang2017chestx}, which achieved state-of-the-art performance on over 14 disease classes and exceeded radiologist performance on pneumonia using the F1 metric. Rajpurkar et \textit{al}.~\cite{rajpurkar2018deep} subsequently developed CheXNeXt, an improved version of the CheXNet, whose performance is on par with radiologists on a total of 10 pathologies of ChestX-ray14. Yao et \textit{al}. proposed~\cite{yao2017learning} to combine a CNN encoder with a Recurrent Neural Network (RNN) decoder to learn not only the visual features of the CXRs in ChestX-ray14 but also the dependencies between their labels. Another notable work based on ChestX-ray14 was by Kumar et \textit{al}.~\cite{kumar2018boosted} who presented a cascaded deep neural network to improve the performance of the multi-label classification task. Closely related to our paper is the work of Chen et \textit{al}.~\cite{Chen2019DeepHM}, in which they proposed to use the \emph{conditional training} strategy to exploit the hierarchy of lung abnormalities in the PLCO dataset~\cite{GOHAGAN2000251S}. In this method, a DenseNet-121 was first trained on a restricted subset of the data such that all parent nodes in the label hierarchy are positive and then finetuned on the whole data.

Recently, the availability of very large-scale CXR datasets such as CheXpert \cite{Irvin2019CheXpertAL} and MIMIC-CXR~\cite{johnson2019mimic} provides researchers with an ideal volume of data  (224,316 scans of CheXpert and more than 350,000 of MIMIC-CXR) for developing better and more robust supervised learning algorithms. Both of these datasets were automatically labeled by the same report-mining tool with 14 common findings. Irvin et \textit{al}.~\cite{Irvin2019CheXpertAL} proposed to train a 121-layer DenseNet on CheXpert with various approaches for handling the uncertainty labels. In particular, uncertainty labels were either ignored (\texttt{U-Ignore} approach) or mapped to \emph{positive} (\texttt{U-Ones} approach) or \emph{negative} (\texttt{U-Zeros} approach). On average, this baseline model outperformed 1.8 out of 3 individual radiologists with an AUC of 0.907 when predicting 5 selected pathologies on a test set of 500 studies. In another work, Rubin et \textit{al}.~\cite{rubin2018large} introduced DualNet---a novel dual convolutional networks that were jointly trained on both the frontal and lateral CXRs of MIMIC-CXR. Experiments showed that the DualNet provides an improved performance in classifying findings in CXR images when compared to separate baseline (\textit{i.e.} frontal and lateral) classifiers. 



In this paper, we adapt the conditional training approach of~\cite{Chen2019DeepHM} to extensively train a series of CNN architectures for the hierarchy of the 14 CheXpert pathologies, which is totally different from that of PLCO. Our approach is significantly different from~\cite{yao2017learning} as we directly exploit a predefined hierarchy of labels instead of learning it from data. Furthermore, unlike previous studies~\cite{Chen2019DeepHM,yao2017learning}, we also propose the use of the label smoothing regularization (LSR)~\cite{Muller2019WhenDL} to leverage uncertainty labels, which, as experiments will later show, significantly improves 
the uncertainty approaches originally proposed in~\cite{Irvin2019CheXpertAL}. 
\section{Proposed Method}
\label{sec:3}
\label{sec:method}
In this section, we present details of the proposed method. We first give a formulation of the multi-label classification for CXRs and the evaluation protocol used in this study (Section \ref{setc.3.1}). We then describe a new training procedure that exploits the relationship among diseases for improving model performance (Section \ref{setc.3.2}). This section also introduces the way we use LSR to deal with uncertain samples in the training data (Section \ref{setc.3.3}).
\subsection{Problem formulation}
\label{setc.3.1}
Our focus in this paper is to develop and evaluate a deep learning-based approach that could learn from hundreds of thousands of CXR images and make accurate diagnoses of 14 common thoracic diseases and observations from unseen samples. These categories include \textit{Enlarged Cardiomediastinum}, \textit{Cardiomegaly}, \textit{Lung Opacity}, \textit{Lung Lesion}, \textit{Edema}, \textit{Consolidation}, \textit{Pneumonia}, \textit{Atelectasis}, \textit{Pneumothorax}, \textit{Pleural Effusion}, \textit{Pleural Other}, \textit{Fracture}, \textit{Support Devices},  and \textit{No Finding}. In this multi-label learning scenario, we are given a training set $\mathcal{D}= \left\{\left(\textbf{x}^{(i)}, \textbf{y}^{(i)}\right); i = 1, \ldots, N\right\}$ that contains $N$ CXRs. A significant portion of the dataset goes with uncertainty labels. It means that each input image $\textbf{x}^{(i)}$ is associated with label $\textbf{y}^{(i)} \in \{0,1,-1\}^{14}$, where 0, 1, and $-1$ correspond to \emph{negative}, \emph{positive}, and \emph{uncertain}, respectively. Note that during the training stage, we apply various approaches to replace all uncertainty labels with \emph{positive}, \emph{negative}, or a smoothed version of one of these two classes. Meanwhile, the  output of the model is a vector of 14 entries, each of which reflects the probability of a specific category being positive. Specifically, we train a CNN, parameterized by weights $\btheta$, that maps $\textbf{x}^{(i)}$  to a prediction $\hat{\textbf{y}}^{(i)} \in [0, 1]^{14}$ such that the cross-entropy loss function is minimized over the training set $\mathcal{D}$. Note that, instead of the softmax function, in the multi-label classification, the sigmoid activation function
\begin{align}
    \hat{y}_k = \dfrac{1}{1 + \exp(-z_k)},\quad k=1,\ldots,14,
\end{align}
is applied to the logits $z_k$ at the last layer of the CNN in order to output each of the 14 probabilities. The loss function is then given by
\begin{align}
    \ell(\btheta) = \sum_{i=1}^{N}\sum_{k=1}^{14} y_k^{(i)}\log\hat{y}_k^{(i)} + \left(1-y_k^{(i)}\right) \log \left(1-\hat{y}_k^{(i)}\right).
\end{align}

\begin{figure}[ht]
  \centering
  \includegraphics[width=8.5cm,height=6.5cm]{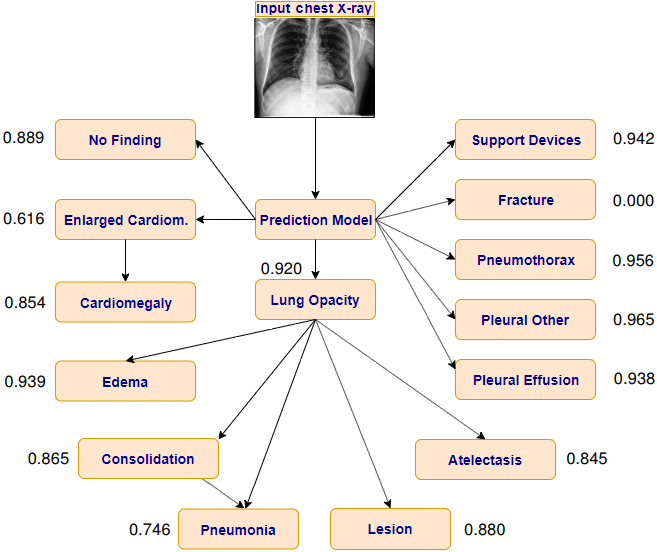}
  \caption{Illustration of our classification task, which aims to build a deep learning system for predicting probability of presence of 14 different pathologies or observations from the CXRs. The relationships among labels were proposed by Irvin et \textit{al}. \cite{Irvin2019CheXpertAL}.}
  \label{fig:1}
\end{figure}

A validation set $\mathcal{V} = \left\{\left(\textbf{x}^{(j)},\textbf{y}^{(j)}\right); j = 1,\ldots, M\right\}$ contains $M$ CXRs, annotated by a panel of 5 radiologists, is used to evaluate the effectiveness of the proposed method. More specifically, model performance is measured by the AUC scores over 5 observations: \textit{Atelectasis}, \textit{Cardiomegaly}, \textit{Consolidation}, \textit{Edema}, and \textit{Pleural Effusion} from the validation set of the CheXpert dataset \cite{Irvin2019CheXpertAL}, which were selected based on clinical importance and prevalence. Figure \ref{fig:1} shows an illustration of the task we investigate in this paper.
\subsection{Conditional training to learn dependencies among labels}
\label{setc.3.2}
In medical imaging, labels are often organized into hierarchies in form of a tree or a directed acyclic graph (DAG). These hierarchies are constructed by domain experts, \textit{e.g.} radiologists in the case of CXR data. Diagnoses or observations in CXRs are often conditioned upon their parent labels~\cite{van2012relationship}. This important fact should be leveraged during the model training and prediction. Most existing CXR classification approaches, however, treat each label in an independent manner and do not take the label structure into account. This group of algorithms is known as \textit{flat classification methods}~\cite{alaydie2012exploiting}. A flat learning model reveals some limitations when applied to hierarchical data as it fails to model the dependency between diseases. For example, from Figure~\ref{fig:1}, the presence of \textit{Cardiomegaly} implies the presence of \textit{Enlarged Cardiomediastinum}. Additionally, some labels that are at the lower levels in the hierarchy, in particular at leaf nodes, have very few positive samples, which makes the flat learning model easily biased toward the negative class.

Another group of algorithms called \textit{hierarchical multi-label classification methods} has been proposed for leveraging the hierarchical relationships among labels in making predictions, which has been successfully exploited for text processing~\cite{aly-etal-2019-hierarchical}, visual recognition~\cite{bi2012mandatory,yan2015hd} and genomic analysis~\cite{bi2015bayes}. The hierarchies are constructed in a way that the root nodes correspond to the most general classes (like  \textit{Opacity}) and the leaf nodes correspond to the most specific ones (like  \textit{Pneumonia}). 
 One common approach~\cite{Chen2019DeepHM} to exploit such a hierarchy is to (1) train a classifier on conditional data, ignoring all samples with negative parent-level labels, and then (2) add these samples back to finetune the network on the whole dataset. Importantly, this strategy is not applied to the validation set since the classifier has been trained on the full dataset during the second phase. Instead, unconditional probabilities should be computed during the inference stage.

\begin{figure}[ht]
  \centering
  \includegraphics[width=12cm,height=3cm]{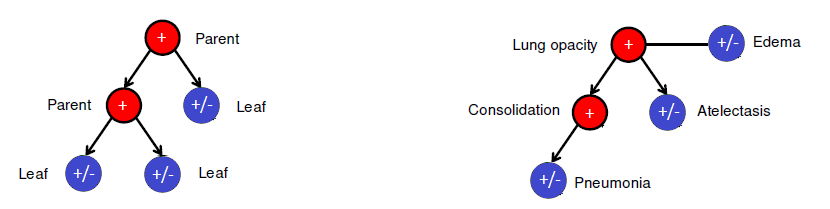}
  \caption{Illustration of the key idea behind the conditional training (\textbf{left}). In this stage, a CNN is trained on a training set where all parent labels (red nodes) are positive, to classify leaf labels (blue nodes), which could be either positive or negative. For example, we train a CNN to classify \textit{Edema}, \textit{Atelectasis}, and \textit{Pneumonia} on training examples where both \textit{Lung Opacity} and \textit{Consolidation} are positive (\textbf{right}).}
  \label{fig:conditional_training}
\end{figure}

We adapt the idea of Chen \textit{et al.}~\cite{Chen2019DeepHM} to the lung disease hierarchy in Figure~\ref{fig:1}, which was initially introduced in~\cite{Irvin2019CheXpertAL}. Presuming the medical validity of the hierarchy, we break the training procedure into two steps. The first step, called \textit{conditional training}, aims to learn the dependent relationships between parent and child labels and to concentrate on distinguishing lower-level labels, in particular the leaf labels. In this step, a CNN is pretrained on a partial training set containing all positive parent labels to classify the child labels; this procedure is illustrated in Figure \ref{fig:conditional_training}.
In the second step, \emph{transfer learning} will be exploited. Specifically, we freeze all the layers of the pretrained network except the last fully connected layer and then retrain it on the full dataset. This training stage aims at improving the capacity of the network in predicting parent-level labels, which could also be either positive or negative. 

\begin{figure}
  \centering
  \includegraphics[width=8.3cm,height=2.2cm]{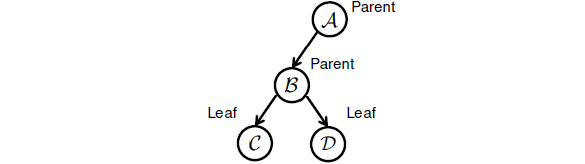}
  \caption{An example of a tree of 4 diseases: $A$, $B$, $C$, and $D$.}
  \label{fig:4}
\end{figure}

According to the above training strategy, the output of the network for each label can be viewed as the conditional probability that this label is positive given its parent being positive. During the inference phase, however, all the labels should be unconditionally predicted. Thus, as a simple application of the Bayes rule, the unconditional probability of each label being positive should be computed by multiplying all conditional probabilities produced by the CNN along the path from the root node to the current label. For illustration, assume a tree of 4 diseases $A, B, C$, and $D$ as shown in Figure~\ref{fig:4} and let $\mathcal{A}$, $\mathcal{B}$, $\mathcal{C}$, and $\mathcal{D}$ be the corresponding events that these labels are positive. Suppose the tuple of conditional predictions $\left(p(\mathcal{A}),p(\mathcal{B}|\mathcal{A}), p(\mathcal{C}|\mathcal{B},\mathcal{A}),p(\mathcal{D}|\mathcal{B},\mathcal{A})\right)$ are already provided by the network. Note that a child label being positive implies that all of its parent labels being positive too. Thus, the unconditional probability of a leaf-node label being positive is identical to the  probability that all labels along the  path from the leaf node tracing back to the root node are jointly positive. In particular, the unconditional predictions for the presence of {$C$ can be computed by}
\begin{align}
   p(\mathcal{C}) &= p(\mathcal{C,B,A}) \\
   &= p(\mathcal{A})p(\mathcal{B} |\mathcal{A})p(\mathcal{C} | \mathcal{\mathcal{B}},\mathcal{A}).
\end{align}
Similarly,
 \begin{align}   
    p(\mathcal{D})&=p(\mathcal{D,B,A}) \\
    &= p(\mathcal{A})p(\mathcal{B} |\mathcal{A})p(\mathcal{D} | \mathcal{B},\mathcal{A}).
 \end{align}

It is important to note that the \emph{unconditional inference} mentioned above helps ensure that the probability of presence of a child disease is always smaller than the probability of its parent, which is consistent with clinical taxonomies in practice.
\subsection{Leveraging uncertainty in CXRs with label smoothing regularization}
\label{setc.3.3}
Another challenging issue in the multi-label classification of CXRs is that we may not have full access to the true labels for all input images provided by the training dataset. A considerable effort has been devoted to creating large-scale CXR datasets with more reliable ground truth, such as CheXpert~\cite{Irvin2019CheXpertAL} and MIMIC-CXR~\cite{johnson2019mimic}. The labeling of these datasets, however, heavily depends on expert systems (\textit{i.e.} using keyword matching with hard-coded rules), which left many CXR images with uncertainty labels. This is mainly due to the unavoidable ambiguities in medical reports.
Several approaches have been proposed in~\cite{Irvin2019CheXpertAL} to deal with these uncertain samples. For example, they can be all \textit{ignored} (\texttt{U-Ignore}), all mapped to \emph{positive}  (\texttt{U-Ones}), or all mapped to \emph{negative} (\texttt{U-Zeros}). 
While \texttt{U-Ignore} could not make use of the full list of labels on the whole dataset, both the \texttt{U-Ones} and \texttt{U-Zeros}  yielded a minimal improvement on CheXpert, as reported in~\cite{Irvin2019CheXpertAL}.
This may be because setting all uncertainty labels to either 1 or 0 will certainly produce a lot of wrong labels, which misguide the model training. 
%
%
%

In this paper, we propose to apply a new advance in machine learning called \emph{label smoothing regularization} (LSR)~\cite{szegedy2016rethinking,pereyra2017regularizing} for a better handling of uncertainty samples. The method has been effectively used~\cite{Muller2019WhenDL} to boost the performances of multi-class classification models via smoothing out the label vector of each sample. We adapt this idea of LSR to the binary classification of a CXR into positive/negative for each of the 14 categories. Our main goal is to exploit the large amount of uncertain CXRs and, at the same time, to prevent the model from overconfident prediction of the training examples that might contain mislabeled data. Specifically, the \texttt{U-ones} approach is softened by mapping each uncertainty label ($-1$) to a random number close to 1. The proposed \texttt{U-ones+LSR} approach now maps the original label $y_k^{(i)}$ to 
\begin{align}
    \bar{y}_k^{(i)} = 
    \begin{cases}
        u, & \text{if } y_k^{(i)}=-1\\ 
        y_k^{(i)} , & \text{otherwise},
    \end{cases}
\end{align}
where $u\sim U(a_1,b_1)$ is a uniformly distributed random variable between $a_1$ and $b_1$---the hyper-parameters of this approach. Similarly, we propose the \texttt{U-zeros+LSR} approach that softens the \texttt{U-zeros} by setting each uncertainty label to a random number $u\sim U(a_0,b_0)$ that is closed to 0.

\section{Experiments} 
\label{sec:4}
\subsection{CXR dataset and settings}
CheXpert dataset~\cite{Irvin2019CheXpertAL} was used to develop and evaluate the proposed method. This is one of the largest public CXR dataset currently available, which contains 224,316 X-ray scans of 65,240 patients. The dataset was labeled for the presence of 14 observations, including 12 common thoracic pathologies. Each observation can be assigned to either \textit{positive} (1), \textit{negative} (0), or \textit{uncertain} (-1).
The main task on  CheXpert is to predict the probability of multiple observations from an input CXR. The predictive models take as input a single view CXR and output the probability of each of the 14 observations as shown in Figure \ref{fig:1}. The whole dataset is divided into a training set of 223,414 studies, a validation set of 200 studies, and a test set of 500 studies. For the validation set, the ground-truth label of each study is obtained by taking the majority vote amongst the annotations of 3 board-certified radiologists. Meanwhile, each study in the test set is labeled by the consensus of 5 board-certified radiologists. The authors of CheXpert proposed an evaluation protocol over 5 observations: \textit{Atelectasis}, \textit{Cardiomegaly}, \textit{Consolidation}, \textit{Edema}, and \textit{Pleural Effusion}, which were selected based on the clinical importance and prevalence from the validation set. The effectiveness of predictive models is measured by the AUC metric. 

\subsection{Data cleaning and normalization}

The learning performance of deep neural networks on raw CXRs may be affected by the irrelevant noisy areas such as texts or the existence of irregular borders. Moreover, we observe a high ratio of CXRs that have poor alignment. We therefore propose a series of preprocessing steps to reduce the effect of irrelevant factors and focus on the lung area. Specifically, all CXRs were first rescaled to $256\times 256$ pixels. A template matching algorithm~\cite{brunelli2009template} was then used to search and find the location of a template chest image ($224 \times 224$ pixels) in the original images. Finally, they were normalized using mean and standard deviation of images from the ImageNet training set~\cite{NIPS2012_4824} in order to reduce source-dependent variation.  

\subsection{Network architecture and training methodology}
The conditional training was performed after applying different approaches for uncertainty labels (\textit{i.e.}  \texttt{U-Ignore}, \texttt{U-Ones}, \texttt{U-Zeros}, \texttt{U-Zeros+LSR}, \texttt{U-Ones+LSR}).
We used DenseNet-121~\cite{huang2017densely} as a baseline network architecture for verifying our hypotheses on the conditional training procedure (Section~\ref{setc.3.2}) and the effect of LSR (Section~\ref{setc.3.3}). In the training stage, all images were fed into the network with a standard size of $224 \times 224$ pixels. The final fully-connected layer is a 14-dimensional dense layer, followed by sigmoid activations that were applied to each of the outputs to obtain the predicted probabilities of the presence of the 14 pathology classes. We used Adam optimizer~\cite{kingma2014adam} with default parameters $\beta_1$ = 0.9, $\beta_2$ = 0.999 and a batch size of 32 to find the optimal weights. The learning rate was initially set to $1e-4$ and then reduced by a factor of 10 after each epoch during the training phase. The network was initialized with a pretrained model on ImageNet~\cite{deng2009imagenet} and then trained for 5 epochs on the conditional data that excludes all examples containing negative parent labels. Next, we added back these samples to the training set and trained the network for 5 more epochs on the full data. During training, our goal is to minimize the binary cross-entropy loss function between the ground-truth labels and the predicted labels output by the network over the training samples. The proposed deep network was implemented in Python using Keras with TensorFlow as backend. All experiments were conducted on a Windows 10 machine with a single NVIDIA Geforce RTX 2080 Ti with 11GB memory.

We conducted extensive ablation studies to verify the impact of the proposed conditional training procedure and LSR. Specifically, we first independently trained  the baseline network with 3 label approaches: \texttt{U-Ignore}, \texttt{U-Ones}, and \texttt{U-Zeros}. We then fixed the hyper-parameter settings of these runs and performed the conditional training procedure on top of them, resulting in 3 other networks: \texttt{U-Ignore+CT}, \texttt{U-Ones+CT}, and \texttt{U-Zeros+CT}, respectively. Next, the LSR technique was applied to the two label approaches \texttt{U-Ones} and \texttt{U-Zeros}. For \texttt{U-Ones}, all uncertainty labels were mapped to random numbers uniformly distributed in the interval $[0.55,\ 0.85]$. For \texttt{U-Zeros}, we labeled uncertain samples with random numbers in $[0,\ 0.3]$. Both of these intervals were emperically chosen. Finally, both CT and LSR were combined with \texttt{U-Ones} and \texttt{U-Zeros} using the same set of hyperparameters, resulting in \texttt{U-Ones+CT+LSR} and \texttt{U-Zeros+CT+LSR}, respectively. Note that all of the above experiments were performed with a template matching (TM) algorithm as a preprocessing step. To isolate the effect of TM, we ran an additional experiment for the baseline \texttt{U-Ignore} with TM being removed.

\subsection{Model ensembling}
In a multi-label classification setting, it is hard for a single CNN model to obtain high and consistent AUC scores for all disease labels. In fact, the AUC score for each label often varies with the choice of network architecture. In order to achieve a highly accurate classifier, an ensemble technique should be explored. The key idea of the ensembling  is to rely on the diversity of a set of possibly weak classifiers that can be combined into a stronger classifier. To that end, we trained and evaluated a strong set of different state-of-the-art CNN models on the CheXpert. The following six architectures were investigated: DenseNet-121, DenseNet-169, DenseNet-201~\cite{huang2017densely}, Inception-ResNet-v2~\cite{szegedy2017inception}, Xception~\cite{chollet2017xception}, and NASNetLarge~\cite{zoph2018learning}. The ensemble model was simply obtained by averaging the outputs of all trained networks. In the inference stage, the test-time augmentation (TTA)~\cite{simonyan2014very} was also applied. Specifically, for each test CXR, we applied a random transformation (amongst horizontal flipping, rotating $\pm$7 degrees, scaling $\pm$2{\footnotesize\%}, and shearing $\pm$5 pixels) 10 times and then averaged the outputs of the model on the 10 transformed samples to get the final prediction. We also carried out the ensembling without TTA to identify the main source of improvement.

\subsection{Quantitative results}\label{sec:result}

\begin{figure}[ht]
    \centering
    \includegraphics[width=3.8cm,height=3.8cm]{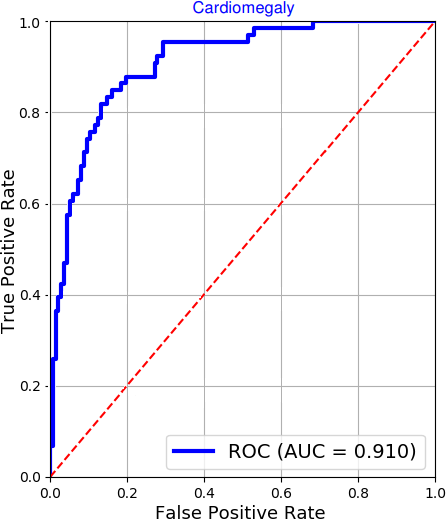}
    \includegraphics[width=3.8cm,height=3.8cm]{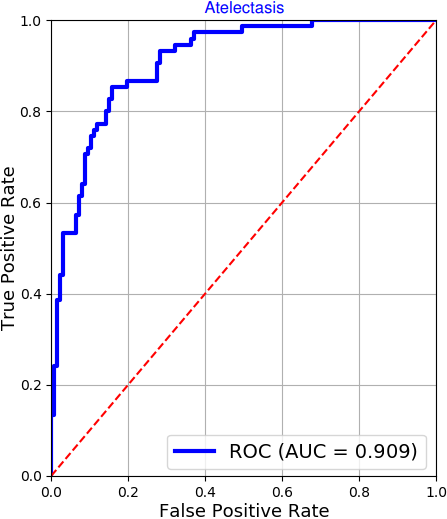}
    \includegraphics[width=3.8cm,height=3.8cm]{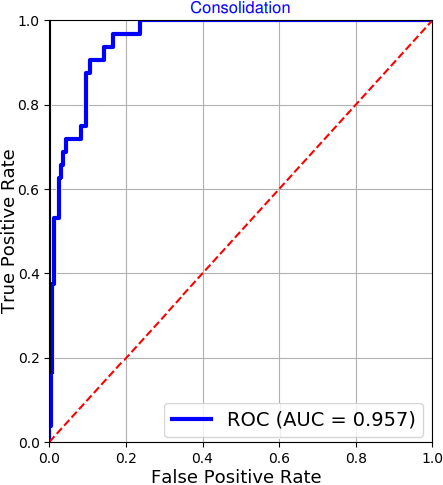}\\[0.2cm]
    \includegraphics[width=3.8cm,height=3.8cm]{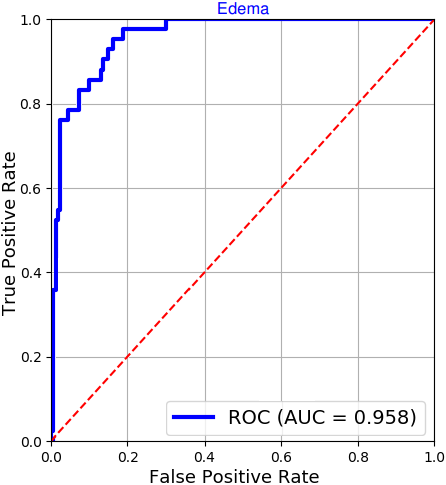}
    \includegraphics[width=3.8cm,height=3.8cm]{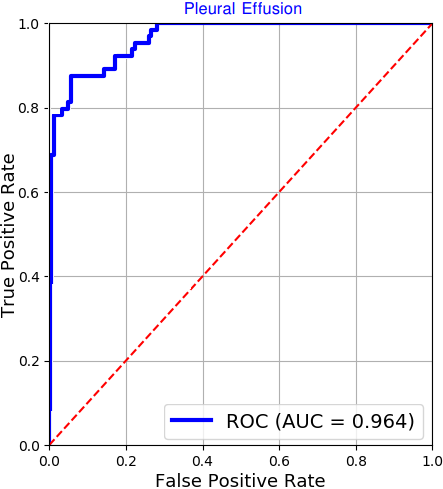}\\[-0.2cm]

    \caption{ROC curves of our ensemble model for the 5 pathologies on CheXpert validation set.}
    \label{fig:roc}
\end{figure}

\begin{figure}[ht]
  \centering
  \includegraphics[width=12cm,height=4.5cm]{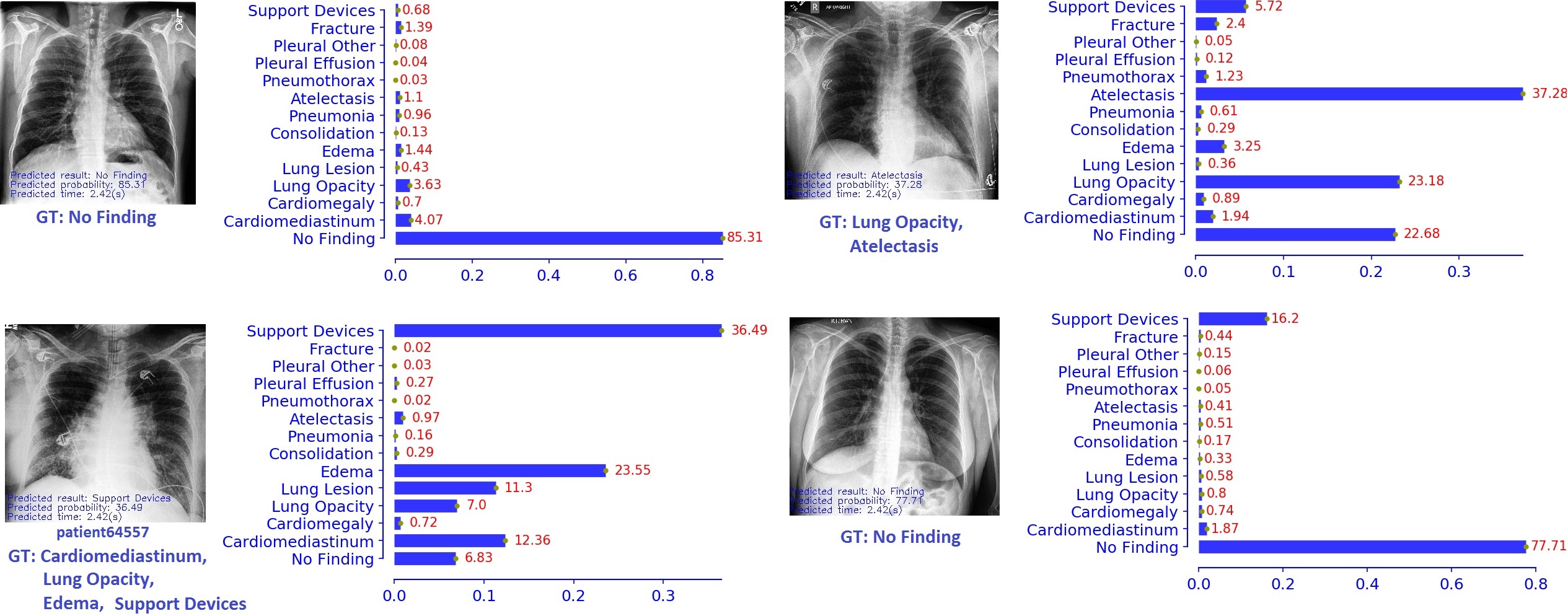}
  \caption{Visualization of findings by the proposed network during the inference stage.}
  \label{fig:predictions}
\end{figure}

Table \ref{tab:2} provides the AUC scores for all experimental settings we have conducted on the CheXpert validation set. We found that the best performing DenseNet-121 model was trained with the \texttt{U-Ones+CT+LSR} approach, which obtained an AUC of 0.894 on the validation set. This is a 4\% improvement compared to the baseline trained with the \texttt{U-Ones} approach (mean AUC = 0.860). Additionally, experimental results show that both the proposed conditional training and LSR help boost the model performance. 
%
Our final model, which is an ensemble of six single models, achieved an average AUC of 0.940. As shown in Table \ref{tab:performance_comparison}, this score outperforms all previous state-of-the-art results. Figure \ref{fig:roc} plots the ROC curves of the ensemble model for 5 pathologies on the validation set. Figure \ref{fig:predictions} illustrates some example predictions by the model during the inference stage.

The effects of using TM and TTA can be seen in Tables~\ref{tab:A3} and~\ref{tab:A4}. While TM improves the mean AUC of DenseNet-121 (with \texttt{U-Ignore} approach) by 0.006, removing TTA from the model ensembling only decreases the mean AUC by 0.003. These gaps are marginal and, so, empirically confirm that the main source of improvement over previous methods indeed comes from our use of conditional training and LSR. 

\begin{table}
\centering
\caption{\label{tab:2} Experimental results on the CheXpert dataset measured by AUC metric over 200 chest radiographic studies of the validation set. \texttt{CT} and \texttt{LSR} stand for \textit{conditional training} and \textit{label smoothing regularization}, respectively. For each label approach, the highest AUC scores are boldfaced.}
\begin{tabular}{p{2cm}p{1.2cm}p{1.35cm}p{1.35cm}p{1.2cm}p{1.25cm}p{1cm}}\\[-0.5cm]
\hline 
\hline 
{\tiny \textbf{Method}}  &  {\tiny \textbf{Atelectasis}} &{\tiny\textbf{Cardiomegaly}}&{\tiny\textbf{Consolidation}} &  {\tiny \hspace*{0.2cm} \textbf{Edema}}  &  {\tiny \textbf{P. Effusion}}   &   {\tiny \hspace*{0.1cm} \textbf{Mean}} \\
\hline                                                         
{\scriptsize  \texttt{U-Ignore}}  &    \hspace*{0.3cm}{\scriptsize 0.768} & \hspace*{0.30cm}{\scriptsize 0.795}  &   \hspace*{0.40cm}{\scriptsize 0.915} &   \hspace*{0.4cm}{\scriptsize \textbf{0.914}} &   \hspace*{0.40cm}{\scriptsize 0.925} &  \hspace{0.2cm}{\scriptsize  0.863}
\\
{\scriptsize \textbf{\texttt{U-Ignore+CT}}} &   \hspace*{0.3cm}{\scriptsize \textbf{0.780}}  &    \hspace*{0.3cm}{\scriptsize \textbf{0.815}} &   \hspace*{0.40cm}{\scriptsize \textbf{0.922}}  &  \hspace*{0.40cm}{\scriptsize \textbf{0.914}}   &  \hspace*{0.40cm}{\scriptsize \textbf{0.928}} &  \hspace{0.2cm}{\scriptsize \textbf{0.872}}\\
\hline
{\scriptsize  \texttt{U-Zeros}}   \cellcolor{gray!30} & \cellcolor{gray!30} \hspace*{0.3cm}{\scriptsize 0.745}  &   \cellcolor{gray!30} \hspace*{0.3cm}{\scriptsize 0.813} & \cellcolor{gray!30} \hspace*{0.40cm}{\scriptsize 0.882} &  \cellcolor{gray!30} \hspace*{0.40cm}{\scriptsize 0.921}   &  \cellcolor{gray!30} \hspace*{0.40cm}{\scriptsize \textbf{0.930}} &  \cellcolor{gray!30} \hspace{0.2cm}{\scriptsize 0.858}\\
{\scriptsize \cellcolor{gray!30} \texttt{U-Zeros+CT}} &   \cellcolor{gray!30} \hspace*{0.3cm}{\scriptsize 0.782}  &    \cellcolor{gray!30} \hspace*{0.3cm}{\scriptsize \textbf{0.835}} &  \cellcolor{gray!30}  \hspace*{0.40cm}{\scriptsize 0.922}  &  \cellcolor{gray!30} \hspace*{0.40cm}{\scriptsize 0.923}   &  \cellcolor{gray!30} \hspace*{0.40cm}{\scriptsize 0.921} &  \cellcolor{gray!30} \hspace{0.2cm}{\scriptsize 0.877}\\
\cellcolor{gray!30} {\scriptsize \texttt{U-Zeros+LSR}}  & \cellcolor{gray!30} \hspace*{0.3cm}{\scriptsize 0.781} & \cellcolor{gray!30} \hspace*{0.30cm}{\scriptsize 0.815}  & \cellcolor{gray!30} \hspace*{0.40cm}{\scriptsize 0.920} &  \cellcolor{gray!30} \hspace*{0.4cm}{\scriptsize 0.923} &  \cellcolor{gray!30} \hspace*{0.40cm}{\scriptsize 0.918} & \cellcolor{gray!30} \hspace*{0.2cm}{\scriptsize 0.871} \\
\cellcolor{gray!30} {\scriptsize \texttt{\textbf{U-Zeros+CT+LSR}}}  & \cellcolor{gray!30} \hspace*{0.3cm}{\scriptsize \textbf{0.806}} & \cellcolor{gray!30} \hspace*{0.30cm}{\scriptsize 0.833}  &  \cellcolor{gray!30} \hspace*{0.40cm}{\scriptsize \textbf{0.929}} &  \cellcolor{gray!30} \hspace*{0.4cm}{\scriptsize \textbf{0.933}} &  \cellcolor{gray!30} \hspace*{0.40cm}{\scriptsize 0.921} &  \cellcolor{gray!30} \hspace*{0.2cm}{\scriptsize \textbf{0.884}} \\
\hline
{\scriptsize \cellcolor{gray!60} \texttt{U-Ones}} & \cellcolor{gray!60}  \hspace*{0.3cm}{\scriptsize 0.800}  &   \cellcolor{gray!60}  \hspace*{0.3cm}{\scriptsize 0.780} & \cellcolor{gray!60}  \hspace*{0.40cm}{\scriptsize 0.882}  & \cellcolor{gray!60}  \hspace*{0.40cm}{\scriptsize 0.918}   & \cellcolor{gray!60}  \hspace*{0.40cm}{\scriptsize 0.920} & \cellcolor{gray!60}  \hspace{0.2cm}{\scriptsize 0.860}\\

{\scriptsize \texttt{U-Ones+CT}} \cellcolor{gray!60} &  \cellcolor{gray!60} \hspace*{0.3cm}{\scriptsize 0.813}  &   \cellcolor{gray!60} \hspace*{0.3cm}{\scriptsize 0.816} &  \cellcolor{gray!60} \hspace*{0.40cm}{\scriptsize 0.895}  &   \cellcolor{gray!60}\hspace*{0.40cm}{\scriptsize 0.923}   &  \hspace*{0.40cm}{\scriptsize 0.912} \cellcolor{gray!60} &   \cellcolor{gray!60}\hspace{0.2cm}{\scriptsize 0.872}\\
{\scriptsize \texttt{U-Ones+LSR}}  \cellcolor{gray!60} &  \cellcolor{gray!60} \hspace*{0.3cm}{\scriptsize 0.818} & \cellcolor{gray!60} \hspace*{0.30cm}{\scriptsize 0.834}  &  \cellcolor{gray!60} \hspace*{0.40cm}{\scriptsize 0.874} &  \cellcolor{gray!60} \hspace*{0.4cm}{\scriptsize 0.925} &  \cellcolor{gray!60} \hspace*{0.40cm}{\scriptsize 0.921} & \cellcolor{gray!60} \hspace*{0.2cm}{\scriptsize 0.874} \\

{\scriptsize \texttt{U-Ones+CT+LSR}}  \cellcolor{gray!60} &  \cellcolor{gray!60} \hspace*{0.3cm}{\scriptsize \textbf{0.825}} &  \cellcolor{gray!60} \hspace*{0.30cm}{\scriptsize \textbf{0.855}}  & \cellcolor{gray!60} \hspace*{0.40cm}{\scriptsize \textbf{0.937}} &  \cellcolor{gray!60} \hspace*{0.4cm}{\scriptsize \textbf{0.930}} &  \cellcolor{gray!60} \hspace*{0.40cm}{\scriptsize \textbf{0.923}} &  \cellcolor{gray!60} \hspace*{0.2cm}{\scriptsize \textbf{0.894}} \\
\hline 
\hline
\end{tabular}
\end{table} 

\begin{table}
\centering
\caption{\label{tab:performance_comparison} Performance comparison using AUC metric between our ensemble of 6 models and previous works on the CheXpert validation set. The highest AUC scores are boldfaced.}
\begin{tabular}{p{2.25cm}p{1.3cm}p{1.3cm}p{1.2cm}p{1.2cm}p{1.3cm}p{0.9cm}}
\\[-0.5cm]
\hline 
{\tiny \textbf{Method}}  &  {\tiny \textbf{Atelectasis}} &   {\tiny \textbf{Cardiomegaly}} & {\tiny \textbf{Consolidation}} &  {\tiny \hspace*{0.2cm} \textbf{Edema}}  & {\tiny \textbf{P. Effusion}} & {\tiny \textbf{Mean}} \\
\hline     

{\scriptsize \texttt{U-Ignore+LP}~\cite{allaouzi2019novel}} &   \hspace*{0.3cm}{\scriptsize 0.720}  &    \hspace*{0.3cm}{\scriptsize 0.870} &   \hspace*{0.40cm}{\scriptsize 0.770}  &  \hspace*{0.40cm}{\scriptsize 0.870}   &  \hspace*{0.20cm}{\scriptsize 0.900 } &  {\scriptsize 0.826}\\

\cellcolor{gray!30}  {\scriptsize \texttt{U-Ignore+BR}~\cite{allaouzi2019novel}} &  \cellcolor{gray!30}  \hspace*{0.3cm}{\scriptsize 0.720}  &  \cellcolor{gray!30}    \hspace*{0.3cm}{\scriptsize 0.880} & \cellcolor{gray!30}   \hspace*{0.40cm}{\scriptsize 0.770}  &  \cellcolor{gray!30} \hspace*{0.40cm}{\scriptsize 0.870}   &  \cellcolor{gray!30}  \hspace*{0.20cm}{\scriptsize 0.900 } &  \cellcolor{gray!30}  {\scriptsize 0.828}\\

{\scriptsize \texttt{U-Ignore+CC}~\cite{allaouzi2019novel}} &   \hspace*{0.3cm}{\scriptsize 0.700}  &    \hspace*{0.3cm}{\scriptsize 0.870} &   \hspace*{0.40cm}{\scriptsize 0.740}  &  \hspace*{0.40cm}{\scriptsize 0.860}   &  \hspace*{0.20cm}{\scriptsize 0.900 } & {\scriptsize 0.814}\\

{\scriptsize \cellcolor{gray!30} \texttt{U-Ignore}~\cite{Irvin2019CheXpertAL}}  &   \cellcolor{gray!30}  \hspace*{0.3cm}{\scriptsize  0.818} & \cellcolor{gray!30}  \hspace*{0.30cm}{\scriptsize 0.828}  & \cellcolor{gray!30}  \hspace*{0.40cm}{\scriptsize 0.938} & \cellcolor{gray!30}  \hspace*{0.4cm}{\scriptsize 0.934} & \cellcolor{gray!30}  \hspace*{0.20cm}{\scriptsize 0.928} & \cellcolor{gray!30}  {\scriptsize  0.889}
\\

{\scriptsize  \texttt{U-Zeros}~\cite{Irvin2019CheXpertAL}}   & \hspace*{0.3cm}{\scriptsize 0.811}  &   \hspace*{0.3cm}{\scriptsize 0.840} & \hspace*{0.40cm}{\scriptsize 0.932} &  \hspace*{0.40cm}{\scriptsize 0.929}   &  \hspace*{0.20cm}{\scriptsize 0.931} &  {\scriptsize 0.888}\\

{\scriptsize \cellcolor{gray!30} \texttt{U-Ones}~\cite{Irvin2019CheXpertAL}} & \cellcolor{gray!30}  \hspace*{0.3cm}{\scriptsize 0.858}  &   \cellcolor{gray!30}  \hspace*{0.3cm}{\scriptsize 0.832} & \cellcolor{gray!30}  \hspace*{0.40cm}{\scriptsize 0.899}  & \cellcolor{gray!30}  \hspace*{0.40cm}{\scriptsize 0.941}   & \cellcolor{gray!30}  \hspace*{0.20cm}{\scriptsize 0.934} & \cellcolor{gray!30}  {\scriptsize 0.893}\\
         
{\scriptsize \texttt{U-MultiClass}~\cite{Irvin2019CheXpertAL}} &   \hspace*{0.3cm}{\scriptsize 0.821}  &    \hspace*{0.3cm}{\scriptsize 0.854} &   \hspace*{0.40cm}{\scriptsize 0.937}  &  \hspace*{0.40cm}{\scriptsize 0.928}   &  \hspace*{0.20cm}{\scriptsize 0.936 } &  {\scriptsize 0.895}\\
{\scriptsize \cellcolor{gray!30} \texttt{U-SelfTrained}~\cite{Irvin2019CheXpertAL}} &  \cellcolor{gray!30} \hspace*{0.3cm}{\scriptsize 0.833}  &    \cellcolor{gray!30} \hspace*{0.3cm}{\scriptsize 0.831} & \cellcolor{gray!30}  \hspace*{0.40cm}{\scriptsize 0.939}  & \cellcolor{gray!30}  \hspace*{0.40cm}{\scriptsize 0.935}   &  \cellcolor{gray!30} \hspace*{0.20cm}{\scriptsize 0.932 } &  \cellcolor{gray!30} {\scriptsize 0.894}\\

{\scriptsize  Ours} &   \hspace*{0.3cm}{\scriptsize \textbf{0.909}}  &     \hspace*{0.3cm}{\scriptsize \textbf{0.910}} &   \hspace*{0.40cm}{\scriptsize \textbf{0.957}}  &   \hspace*{0.40cm}{\scriptsize \textbf{0.958}}   &   \hspace*{0.20cm}{\scriptsize \textbf{0.964}} &   {\scriptsize \textbf{0.940}}\\

\hline
\end{tabular}
\end{table} 

\begin{table}[ht]
\centering
\caption{\label{tab:A3} AUC improvement on the CheXpert validation set by using TM for the \texttt{U-Ignore} training procedure.\vspace{0.1cm}}
\begin{tabular}{p{2cm}p{1.2cm}p{1.35cm}p{1.35cm}p{1.2cm}p{1.25cm}p{1cm}}
\hline 
{\tiny \textbf{Method}}  &  {\tiny \textbf{Atelectasis}} &{\tiny\textbf{Cardiomegaly}}&{\tiny\textbf{Consolidation}} &  {\tiny \hspace*{0.2cm} \textbf{Edema}}  &  {\tiny \textbf{P. Effusion}}   &   {\tiny \hspace*{0.1cm} \textbf{Mean}} \\
\hline
{\scriptsize  \texttt{U-Ignore}}  &    \hspace*{0.3cm}{\scriptsize 0.776} & \hspace*{0.30cm}{\scriptsize 0.785}  &   \hspace*{0.40cm}{\scriptsize 0.915} &   \hspace*{0.4cm}{\scriptsize 0.913} &   \hspace*{0.40cm}{\scriptsize 0.894} &  \hspace{0.2cm}{\scriptsize  0.857}\\
{\scriptsize  \texttt{U-Ignore+TM}}  &    \hspace*{0.3cm}{\scriptsize 0.768} & \hspace*{0.30cm}{\scriptsize 0.795}  &   \hspace*{0.40cm}{\scriptsize 0.915} &   \hspace*{0.4cm}{\scriptsize 0.914} &   \hspace*{0.40cm}{\scriptsize 0.925} &  \hspace{0.2cm}{\scriptsize  0.863}\\
\hline
\end{tabular}
\end{table}
\begin{table}
\centering
\caption{\label{tab:A4} AUC improvement on the CheXpert validation set by using TTA on top of model ensembling.\vspace{0.1cm}}
\begin{tabular}{p{2.85cm}p{1cm}p{1.3cm}p{1.2cm}p{1.2cm}p{1.3cm}p{0.6cm}}
\hline 
{\tiny \textbf{Method}}  &  {\tiny \textbf{Atelectasis}} &   {\tiny \textbf{Cardiomegaly}} & {\tiny \textbf{Consolidation}} &  {\tiny \hspace*{0.2cm} \textbf{Edema}}  & {\tiny \textbf{P. Effusion}} & {\tiny \textbf{Mean}} \\
\hline   
{\scriptsize  Ensemble with TTA} &   \hspace*{0.3cm}{\scriptsize 0.909}  &     \hspace*{0.3cm}{\scriptsize 0.910} &   \hspace*{0.40cm}{\scriptsize 0.957}  &   \hspace*{0.40cm}{\scriptsize 0.958}   &   \hspace*{0.20cm}{\scriptsize 0.964} &   {\scriptsize 0.940}\\
{\scriptsize  Ensemble without TTA} &   \hspace*{0.3cm}{\scriptsize 0.908}  &     \hspace*{0.3cm}{\scriptsize 0.906} &   \hspace*{0.40cm}{\scriptsize 0.955}  &   \hspace*{0.40cm}{\scriptsize 0.951}   &   \hspace*{0.20cm}{\scriptsize 0.958} &   {\scriptsize 0.937}\\
\hline
\end{tabular}
\end{table} 

\subsection{Independent evaluation and comparison to radiologists}

A crucial evaluation of any machine learning-based medical diagnosis system (ML-MDS) is to evaluate how well the system performs on an independent test set in comparison to human expert-level performance. To this end, we evaluated the proposed method on the hidden test set of CheXpert, which contains 500 CXRs labeled by 8 board-certified radiologists. The annotations of 3 of them were used for benchmarking radiologist performance and the majority vote of the other 5 served as ground truth. For each of the 3 individual radiologists, the AUC scores for the 5 selected diseases (\textit{Atelectasis}, \textit{Cardiomegaly}, \textit{Consolidation}, \textit{Edema}, and \textit{Pleural Effusion}) were computed against the ground truth to evaluate radiologists' performance. We then evaluated our ensemble model on the test set and performed ROC analysis to compare the model performance to radiologists. For more details, the ROCs produced by the prediction model and the three radiologists' operating points were both plotted. For each disease, whether the model is superior to radiologists' performances was determined by counting the number of radiologists' operating points lying below the ROC\footnote{This test was conducted independently with the support of the Stanford Machine Learning Group as the test set is not released to the public.}. The result shows that our deep learning model, when being averaged over the 5 diseases, outperforms 2.6 out of 3 radiologists with an AUC of 0.930. This is the best performance on the CheXpert leaderboard to date. The attained AUC score validates the generalization capability of the trained deep learning model on an unseen dataset. Meanwhile, the total number of radiologists under ROC curves indicates that the proposed method is able to reach human expert-level performance---an important step towards the application of an ML-MDS in real-world scenarios.

\section{Discussions}
\label{sec:5}
\subsection{Key findings and meaning}

By training a set of strong CNNs on a large scale dataset, we built a deep learning model that can accurately predict multiple thoracic diseases from CXRs. In particular, we empirically showed a major improvement, in terms of AUC score, by exploiting the dependencies among diseases and by applying the label smoothing technique to uncertain samples. We found that it is especially difficult to obtain a good AUC score for all diseases with a single CNN. It is also observed that the classification performance varies with network architectures, the rate of positive/negative samples, as well as the visual features of the lung disease being detected. 
In this case, an ensemble of multiple deep learning models plays a key in boosting the generalization of the final model and its performance. Our findings, along with recent publications~\cite{rajpurkar2017chexnet,guan2018diagnose,rajpurkar2018deep,kumar2018boosted}, continue to assert that deep learning algorithms can accurately identify the risk of many thoracic diseases and is able  to assist patient screening, diagnosing, and physician training.

\subsection{Limitations}
Although a highly accurate performance has been achieved, we acknowledge that the proposed method reveals some limitations. The conditional training strategy requires a predefined hierarchy of diseases, which is not easy to construct and usually imperfect. Furthermore, it seems difficult to extend this idea to deeper hierarchies of diseases, for which too many examples have to be excluded from the training set in the first phase. The use of LSR in this paper with heuristically chosen hyper-parameters, while significantly improving the classification performance, is not clearly justified. In addition, the use of TTA has also a limitation due to it decreases inference time.

Other challenges are related to the training data. For instance, the deep learning algorithm was trained and evaluated on a CXR data source collected from a single tertiary care academic institution. Therefore, it may not yield the same level of performance when applied to data from other sources such as from other institutions with different scanners. This phenomenon is called \textit{geographic variation}. To overcome this, the learning algorithm should be trained on data that are diverse in terms of regions, races, imaging protocols, etc. Next, to make a diagnosis from a CXR, doctors often rely on a broad range of additional data such as patient age, gender, medical history, clinical symptoms, and possibly CXRs from different views. This additional information should also be incorporated into the model training. Finally, CXR image quality is another problem. When taking a deeper look at the CheXpert, we found a considerable rate of samples in low quality (\textit{e.g.} rotated image, low-resolution, samples with texts, noise, etc.) that definitely hurts the model performance. In this case, a template matching-based method as proposed in this work may be insufficient to effectively remove all the undesired samples. A more robust preprocessing technique, such as that proposed in~\cite{ccalli2019frodo}, should be applied to reject almost all \textit{out-of-distribution} samples.

\section{Conclusion} 
\label{sec:6}
We presented in this paper a comprehensive approach for building a high-precision computer-aided diagnosis system for common thoracic diseases classification from CXRs. We investigated almost every aspect of the task including data cleaning, network design, training, and ensembling. In particular, we introduced a new training procedure in which dependencies among diseases and uncertainty labels are effectively exploited and integrated in training advanced CNNs. Extensive experiments demonstrated that the proposed method  outperforms the previous state-of-the-art by a large margin on the CheXpert dataset. More importantly, our deep learning algorithm exhibited a performance on par with specialists in an independent test. There are several possible mechanisms to improve the current method. The most promising direction is to increase the size and quality of the dataset. A larger and high-quality labeled dataset can help deep neural networks generalize better and reduce the need for transfer learning from ImageNet. For instance, extra training data from MIMIC-CXR~\cite{johnson2019mimic}, which uses the same labeling tool as CheXpert, should be considered. We are currently expanding this research by collecting a new large-scale CXR dataset with radiologist-labeled reference from several hospitals and medical centers in Vietnam. The new dataset is needed to validate the proposed method and to confirm its usefulness in different clinical settings. 
We believe the cooperation between a machine learning-based medical diagnosis system and radiologists will improve the outcomes of thoracic disease diagnosis and bring benefits to clinicians and their patients.

\section{Acknowledgements} 
This research was supported by the Vingroup Big Data Institute (VinBDI). The authors gratefully acknowledge Jeremy Irvin from the Machine Learning Group, Stanford University for helping us  evaluate the proposed method on the hidden test set of CheXpert.

\bibliography{references}
\

\
\end{document}